\makeatletter
\input{filecontents.sty}
\makeatother

\documentclass[%
titlepage,
charprotruding,
secnumarabic,
showpacs,
twocolumn,
twoside,aps,pre]{itpprepr}

\def\Preprint{}

\makeatletter
\ifx\@undefined\Preprint
  \def\JournalPreprint#1#2{#1}
\else
  \def\JournalPreprint#1#2{#2}
\fi
\makeatother

\JournalPreprint{\documentclass[twocolumn,aps,pre,showpacs,groupedaddress]{revtex4}}{}

\usepackage[american]{babel}
\usepackage{amsfonts}
\usepackage{amsmath}
\usepackage{amssymb}
\usepackage{graphicx}
\usepackage{color}
\usepackage{calc}
\usepackage{ntheorem}
\usepackage[pdftitle={Random numbers for large scale distributed Monte Carlo simulations},
pdfauthor={Heiko Bauke and Stephan Mertens},
pdfcreator={},
pdfsubject={Random Numbers},
pdfproducer={},
pdfkeywords={Monte Carlo, pseudo random numbers, parallel computer},hyperfootnotes=false,breaklinks=true]{hyperref}
\usepackage{url}
\JournalPreprint{}{\usepackage{breakurl}}

\graphicspath{{figures/}}

\IfFileExists{date.tex}{%

}

\newcommand*{\bigO}[1]{\mathcal{O}\left(#1\right)}

\newcommand*{\e}{\mathrm{e}}
\newcommand*{\F}{\mathbb{F}}

\newcommand*{\Z}{\mathbb{Z}}

\newcommand{\bigo}{\ensuremath{\mathcal{O}}}

\newcommand{\definedas}{\mathrel{\raise.095ex\hbox{:}\mkern-4.2mu=}}
\newsavebox{\eqbox}
\sbox{\eqbox}{$\null=\null$}
\newcommand{\Vdots}{\makebox[\wd\eqbox]{\vdots}}

\begin{document}

\theoremstyle{plain}
\JournalPreprint{%
  \theoremheaderfont{\normalfont\bfseries} %
}{%
  \theoremheaderfont{\normalfont\bfseries\sffamily}%
}
\theorembodyfont{\normalfont}
\theoremseparator{}
\setlength{\theoremindent}{0cm}
\setlength{\theorempreskipamount}{0.5\baselineskip plus 0.5em}
\setlength{\theorempostskipamount}{0.5\baselineskip plus 0.5em}
\theoremnumbering{arabic}
\theoremsymbol{}
\newtheorem{theorem}{Theorem}
\newcommand*{\Figurename}{Figure}
\newcommand*{\Tablename}{Table}

\title{Random numbers for large scale distributed Monte Carlo simulations}

\author{Heiko Bauke}
\email[E-mail:]{heiko.bauke@physics.ox.ac.uk}
\affiliation{Institut f\"ur Theoretische Physik, Universit\"at Magdeburg, Universit\"atsplatz~2, 39016~Magdeburg, Germany}
\affiliation{Rudolf Peierls Centre for Theoretical Physics, University of Oxford, 1~Keble~Road, Oxford, OX1~3NP, United Kingdom}
\author{Stephan Mertens}
\email[E-mail:]{stephan.mertens@physik.uni-magdeburg.de}
\affiliation{Institut f\"ur Theoretische Physik, Universit\"at Magdeburg, Universit\"atsplatz~2, 39016~Magdeburg, Germany}
\date{\today}

\begin{abstract}
  Monte Carlo simulations are one of the major tools in statistical physics,
  complex system science, and other fields, and an increasing number of these
  simulations is run on distributed systems like clusters or grids. This
  raises the issue of generating random numbers in a parallel, distributed
  environment. In this contribution we demonstrate that multiple linear
  recurrences in finite fields are an ideal method to produce high quality
  pseudorandom numbers in sequential and parallel algorithms.  Their known
  weakness (failure of sampling points in high dimensions) can be overcome by
  an appropriate delinearization that preserves all desirable properties of
  the underlying linear sequence.
\end{abstract}


\maketitle

\section{Introduction}
\label{sec:intro}

The Monte Carlo method is a major industry and random numbers are its key
resource. In contrast to most commodities, quantity and quality of randomness
have an inverse relation: The more randomness you consume, the better it has
to be \cite{Hayes:2001:Randomness}. The quality issue arises because
simulations only \emph{approximate} randomness by generating a stream of
deterministic numbers, named \emph{pseudo}random numbers (PRNs), with
successive calls to a pseudorandom number generator (PRNG).  Considering the
ever increasing computing power, however, the quality of PRNGs is a moving
target.  Simulations that consume $10^{10}$ pseudorandom numbers were
considered large-scale Monte Carlo simulation a few years ago. On a
present-day desktop machine this is a ten minute run.

More and more large scale simulations are run on parallel systems like
networked workstations, clusters, or ``the grid.''  In a parallel environment
the quality of a PRNG is even more important, to some extent because feasible
sample sizes are easily $10,\ldots, 10^5$ times larger than on a sequential
machine.  The main problem is the parallelization of the PRNG itself, however.
Some good generators are not parallelizable at all, others lose their
efficiency, their quality, or even both when being parallelized
\cite{Coddington:1996:RNG_for_Parallel_Computers,Anderson:1990:RNG}.

In this contribution we discuss linear recurrences in finite fields.  They
excel as sources of pseudorandomness because they have a solid mathematical
foundation and they can easily be parallelized.  Their linear structure,
however, may cause problems in experiments that sample random points in high
dimensional space. This is a known issue (``Random numbers fall mainly in the
planes'' \cite{Marsaglia:1968:planes}), but it can be overcome by a proper
delinearization of the sequence.

The paper is organized as follows. In section \ref{sec:basics} we briefly
review some properties of linear recurrences in finite fields, and we motivate
their use as PRNGs. In section \ref{sec:parallelization} we discuss various
techniques for parallelizing a PRNG. We will see that only two of these
techniques allow for parallel simulations whose results \emph{do not}
necessarily depend on the number of processes, and we will see how linear
recurrences support these parallelization techniques. Section
\ref{sec:delinear} addresses the problem of linear sequences to sample points
in high dimensions, measured by the spectral test. We show that with a proper
transformation (delinearization) we can generate a new sequence that shares
all nice properties with the original sequence (like parallelizability,
equidistribution properties, etc.), but passes many statistical tests that are
sensitive to the hyperplane structure of points of linear sequences in high
dimensions. In the last two sections we discuss the quality of the linear and
nonlinear sequences in parallel Monte Carlo applications and how to implement
these generators efficiently.

\section{Recurrent randomness}
\label{sec:basics}

Almost all PRNGs produce a sequence $(r) = r_1,r_2,\ldots$ of pseudorandom
numbers by a recurrence 
\begin{equation}
  \label{eq:recurrence}
  r_i = f(r_{i-1}, r_{i-2}, \ldots, r_{i-k})\,,
\end{equation}
and the art of random number generation lies in the design of the function
$f$. Numerous recipes for $f$ have been discussed in the literature
\cite{Knuth:1998:TAoCPII,LEcuyer:2004:Random_Number_Generation}. One of the
oldest and most popular PRNGs is the linear congruential generator (LCG)
\begin{equation}
  \label{eq:lehmer}
  r_i = a \cdot r_{i-1} + b \bmod m
\end{equation}
introduced by Lehmer \cite{Lehmer:1951:LCG}. The additive constant $b$ may be
zero.  Knuth \cite{Knuth:1969:TAoCPII} proposed a generalization of Lehmer's
method known as multiple recursive generator (MRG)
\cite{Tezuka:book,LEcuyer:2004:Random_Number_Generation,LEcuyer:Blouin:Couturee:1993:Good_MRGs}
\begin{equation}
  \label{eq:lfsr}
  r_i = a_1\cdot r_{i-1} + a_2\cdot r_{i-2} + \ldots + a_n\cdot r_{i-n} \bmod m\,.
\end{equation}
Surprisingly, on a computer \emph{all} recurrences are essentially linear.
This is due to the simple fact that computers operate on numbers of finite
precision; let us say integers between $0$ and $r_{\mathrm{max}}$. This has
two important consequences:
\begin{itemize}
\item Every recurrence (\ref{eq:recurrence}) must be periodic. 
\item We can embed the sequence $(r)$ into the finite field $\F_m$, the field
  of integers with arithmetic modulo $m$, where $m$ is a prime number larger
  than the largest value in the sequence~$(r)$.
\end{itemize}
The relevance of linear sequences is then based on the following fact
\cite[corollary 6.2.3]{Jungnickel:1993:Finite_Fields}:
\begin{theorem}
  \label{the:linear_recurrences}
  All finite length sequences and all periodic sequences over a finite field
  $\F_m$ can be generated by a linear recurrence \eqref{eq:lfsr}.
\end{theorem}
In the theory of finite fields, a sequence of type (\ref{eq:lfsr}) is called
\emph{linear feedback shift register sequence}, or LFSR sequence for short.
Note that a LFSR sequence is fully determined by specifying $n$ coefficients
$(a_1, a_2, \dots,a_n)$ plus $n$ initial values $(r_1, r_2,\dots,r_n)$.

According to Theorem~\ref{the:linear_recurrences} all finite length sequences
and all periodic sequences are LFSR sequences.  The \emph{linear complexity}
of a sequence is the minimum order of a linear recurrence that generates this
sequence. The Berlekamp-Massey algorithm
\cite{Massey:1969:Shift-register_synthesis,Jungnickel:1993:Finite_Fields} is
an efficient procedure for finding this shortest linear recurrence.  The
linear complexity of a \emph{random} sequence of length $T$ on average equals
$T/2$ \cite{Menezes:1996:Handbook}.  As a consequence a random sequence cannot
be compressed by coding it as a linear recurrence, because $T/2$ coefficients
plus $T/2$ initial values have to be specified. Typically the linear
complexity of a nonlinear sequence (\ref{eq:recurrence}) is of the same order
of magnitude as the period of the sequence.  Since the designers of PRNGs
strive for ``astronomically'' large periods, implementing nonlinear generators
as a linear recurrence is not tractable.  As a matter of principle, however,
all we need to discuss are linear recurrences, and any nonlinear recurrence
can be seen as an efficient implementation of a large order linear
recurrence. The popular PRNG ``Mersenne Twister''
\cite{Matsumoto:Nishimura:1998:Mersenne_Twister}, for example, is an efficient
implementation of a linear recurrence in $\F_2$ of order 19\,937.
 
There is a wealth of rigorous results on LFSR sequences that can (and should)
be used to construct a good PRNG. Here we only discuss a few but important
facts without proofs. A detailed discussion of LFSR sequences including
proofs can be found in
\cite{Golomb:1982:Shift_Register_Sequences,Jungnickel:1993:Finite_Fields,Lidl:Niederreiter:1994:finite_fields,Lidl:Niederreiter:1997:finite_fields,Fillmore:Marx:1968:LRS,Zierler:1959:Linear_Recurring_Sequences}.

Since the all zero tuple $(0,0,\ldots,0)$ is a fixed point of (\ref{eq:lfsr}),
the maximum period of a LFSR sequence cannot exceed \mbox{$m^n-1$}. The
following theorem tells us precisely how to choose the coefficients
$(a_1,a_2,\dots,a_n)$ to achieve this period \cite{Knuth:1998:TAoCPII}.
\begin{theorem}
  \label{the:period}
  The LFSR sequence (\ref{eq:lfsr}) over $\F_m$ has period \mbox{$m^n-1$}, if
  and only if the characteristic polynomial
  \begin{equation}
    \label{eq:feed-poly}
    f(x) = x^n - a_1 x^{n-1} - a_2 x^{n-2} - \ldots - a_n
  \end{equation}
  is \emph{primitive} modulo $m$.
\end{theorem}
A monic polynomial $f(x)$ of degree $n$ over $\F_m$ is primitive modulo $m$,
if and only if it is irreducible (i.\,e., cannot be factorized over $\F_m$),
and if it has a primitive element of the extension field $\F_{m^n}$ as one of
its roots.  The number of primitive polynomials of degree $n$ modulo $m$ is
equal to $\phi(m^n-1)/n=\bigO{m^n/(n\ln (n \ln m))}$
\cite{Wan:2003:Finite_Fields}, where $\phi(x)$ denotes Euler's totient
function. As a consequence a random polynomial of degree $n$ is primitive
modulo $m$ with probability $\simeq 1/(n\ln(n \ln m))$, and finding primitive
polynomials reduces to testing whether a given polynomial is primitive. The
latter can be done efficiently, if the factorization of $m^n-1$ is known
\cite{Jungnickel:1993:Finite_Fields}, and most computer algebra systems offer
a procedure for this.

\begin{theorem}
  \label{the:pn}
  Let $(r)$ be an LFSR sequence (\ref{eq:lfsr}) with a primitive
  characteristic polynomial. Then each $k$-tuple 
  $(r_{i+1}, \ldots, r_{i+k})\in\{0,1,\ldots,m-1\}^k$
  occurs $m^{n-k}$ times per period for $k \leq n$ (except the all zero tuple
  for $k=n$, which never occurs).
\end{theorem}
From this theorem it follows that, if a tuple of $k$ successive terms
$k\leq n$ is drawn from a random position of a LFSR sequence, the
outcome is uniformly distributed over all possible $k$-tuples in
$\F_m$. This is exactly what one would expect from a truly random
sequence.  In terms of Compagner's ensemble theory, tuples of length
$k\leq n$ of a LFSR sequence with primitive characteristic polynomial
are indistinguishable from truly random tuples
\cite{Compagner:1991:Definitions_of_randomness,Compagner:1991:hierarchy_of_correlations}.

\begin{theorem}
  \label{the:cor}
  Let $(r)$ be an LFSR sequence (\ref{eq:lfsr}) with period $T = m^n-1$
  and let $\alpha$ be a complex $m$th root of unity and $\overline{\alpha}$ its
  complex conjugate. Then 
  \begin{equation}
    \label{eq:autocorr}
    C(h) \definedas \sum_{i=1}^{T} \alpha^{r_i}\cdot\overline{\alpha}^{r_{i+h}} =
    \begin{cases}
      T & \text{if $h = 0 \bmod T$} \\
     -1 & \text{if $h \neq 0 \bmod T$.}
    \end{cases}
  \end{equation}
\end{theorem}
$C(h)$ can be interpreted as the autocorrelation function of the sequence, and
Theorem~\ref{the:cor} tells us that LFSR sequences with maximum period have
autocorrelations that are very similar to the autocorrelations of a random
sequence with period $T$. Together with the nice equidistribution properties
(Theorem~\ref{the:pn}) this qualifies LFSR sequences with maximum period as
\emph{pseudonoise sequences}, a term originally coined by Golomb for binary
sequences
\cite{Golomb:1982:Shift_Register_Sequences,Jungnickel:1993:Finite_Fields}.

\section{Parallelization}
\label{sec:parallelization}

\subsection{General parallelization techniques}

\begin{figure}[t]%
  \includegraphics[width=\linewidth]{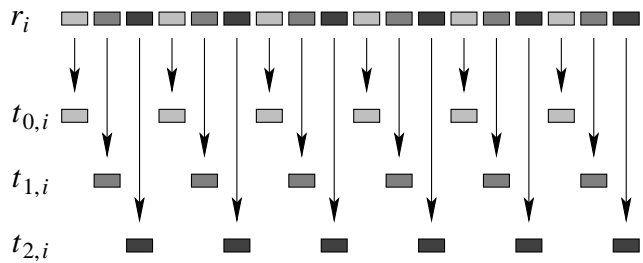}%
  \caption{Parallelization by leapfrogging.}%
  \label{fig:leap_frog}%
\end{figure}

In parallel applications we need to generate streams $t_{j,i}$ of random
numbers, where $j=0,1,\dots,p-1$ numbers the streams for each of the $p$
processes. We require statistical independence of the $t_{j,i}$ within each
stream and between the streams.  Four different parallelization techniques are
used in practice:
\begin{description}
\item[Random seeding:] All processes use the same PRNG but a different
  ``random'' seed. The hope is that they will generate nonoverlapping and
  uncorrelated subsequences of the original PRNG.  This hope, however, has no
  theoretical foundation. Random seeding is a violation of Donald Knuth's
  advice, ``Random number generators should not be chosen at random''
  \cite{Knuth:1998:TAoCPII}.
\item[Parameterization:] All processes use the same type of generator but
  with different parameters for each processor. Example: linear congruential
  generators with additive constant $b_j$ for the $j$th stream
  \cite{Percus:Kalos:1989:RNGs_for_MIMD}:
  \begin{equation}
    t_{j,i}=a\cdot t_{j, i-1} + b_j \bmod m \,,
  \end{equation}
  where the $b_j$'s are different prime numbers just below
  $\sqrt{m/2}$. Another variant uses different multipliers $a$ for different
  streams \cite{Mascagni:1998:Parallel_LCG}.  The theoretical foundation of
  these methods is weak, and empirical tests have revealed serious
  correlations between streams \cite{Matteis:Pagnutti:1990:parallel_PRNGs}.
  On a massive parallel system you may need thousands of parallel streams, and
  it is not trivial to find a type of PRNG with thousands of ``well tested''
  parameter sets.
  \begin{figure}[t]%
    \includegraphics[width=\linewidth]{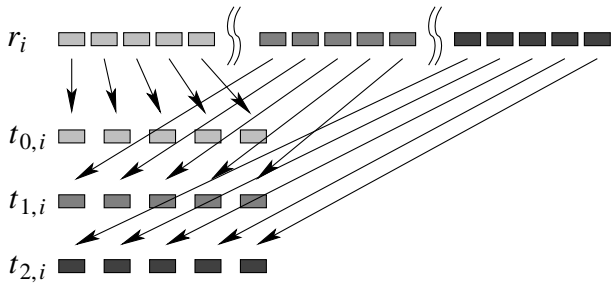}%
    \caption{Parallelization by block splitting.}%
    \label{fig:blocks}%
  \end{figure}
\item[Block splitting:] Let $M$ be the maximum number of calls to a
  PRNG by each processor, and let $p$ be the number of processes.
  Then we can split the sequence $(r)$ of a sequential PRNG into
  consecutive blocks of length $M$ such that
  \begin{equation}
    \begin{split}
      t_{0, i} & = r_{i} \\
      t_{1, i} & = r_{i+M} \\
      & \Vdots\\
      t_{p-1, i} & = r_{i+M(p-1)}\,. 
    \end{split}
  \end{equation}
  This method works only if we know $M$ in advance or can at least safely
  estimate an upper bound for $M$. To apply block splitting it is necessary to
  jump from the $i$th random number to the $(i+M)$th number without
  calculating the numbers in-between, which cannot be done efficiently for
  many PRNGs.  A potential disadvantage of this method is that long range
  correlations, usually not observed in sequential simulations, may become
  short range correlations between substreams
  \cite{Matteis:Pagnutti:1990:Long-range_correlations,Eichenauer-Herrmann:Grothe:1989:long-range_correlations}.
  Block splitting is illustrated in \figurename~\ref{fig:blocks}.
\item[Leapfrog:] The leapfrog method distributes a sequence $(r)$ of random
  numbers over $p$ processes by decimating this base sequence such that
  \begin{equation}
    \label{eq:leap_frog}
    \begin{split}
      t_{0, i} & = r_{pi} \\
      t_{1, i} & = r_{pi+1} \\
      & \Vdots\\
      t_{p-1, i} & = r_{pi+(p-1)}\,. 
    \end{split}
  \end{equation}
  Leapfrogging is illustrated in \figurename~\ref{fig:leap_frog}. It is the
  most versatile and robust method for parallelization and it does not require
  an \emph{a priori} estimate of how many random numbers will be consumed by
  each processor. An efficient implementation requires a PRNG that can be
  modified to generate directly only every $k$th element of the original
  sequence.  Again this excludes many popular PRNGs.
\end{description}
Note that for a periodic sequence $(r)$ the substreams derived from block
splitting are cyclic shifts of the original sequence $(r)$, and for $p$ not
dividing the period of $(r)$, the leapfrog sequences are cyclic shifts of each
other. Hence the leapfrog method is equivalent to block splitting on a
different base sequence.

\subsection{Playing fair}

We say that a parallel Monte Carlo simulation \emph{plays fair}, if its
outcome is strictly independent of the underlying hardware, where strictly
means deterministically, not just statistically. Fair play implies the use of
the same PRNs in the same context, independently of the number of parallel
processes. It is mandatory for debugging, especially in parallel environments
where the number of parallel processes varies from run to run, but another
benefit of playing fair is even more important: Fair play guarantees that the
quality of a PRNG with respect to an application does not depend on the degree
of parallelization.

Obviously, parametrization or random seeding as discussed above prevent a
simulation from playing fair. Leapfrog and block splitting, on the other hand,
do allow one to use the same PRNs within the same context independently of the
number of parallel streams.

Consider the site percolation problem. A site in a lattice of size $N$ is
occupied with some probability, and the occupancy is determined by a PRN. $M$
random configurations are generated.  A naive parallel simulation on $p$
processes could split a base sequence into $p$ leapfrog streams and having
each process generate $\approx M/p$ lattice configurations, independently of
the other processes. Obviously this parallel simulation is not equivalent to
its sequential version that consumes PRNs from the base sequence to generate
one lattice configuration after another. The effective shape of the resulting
lattice configurations depends on the number of processes. This parallel
algorithm does not play fair.

We can turn the site percolation simulation into a fair playing algorithm by
leapfrogging on the level of lattice configurations. Here each process
consumes distinct contiguous blocks of PRNs from the sequence $(r)$, and the
workload is spread over $p$ processors in such a way, that each process
analyzes each $p$th lattice. If we number the processes by their rank $i$ from
$0$ to $p-1$ and the lattices from $0$ to $M-1$, each process starts with a
lattice whose number equals its own rank. That means process $i$ has to skip
$i\cdot N$ PRNs before the first lattice configuration is generated.
Thereafter each process can skip $p-1$ lattices, i.\,e., $(p-1)\cdot N$ PRNs
and continue with the next lattice.

Organizing simulation algorithms such that they play fair is not always as
easy as in the above example, but with a little effort one can achieve fair
play in more complicated situations, too. This may require the combination of
block splitting and the leapfrog method, or iterated leapfrogging.  Sometimes
it is also necessary to use more than one stream of PRNs per process, e.\,g.,
in the Swendsen Wang cluster algorithm one may use one PRNG to construct the
bond percolation clusters and another PRNG to decide to flip the cluster.

\subsection{Parallelization of LFSR sequences}

As a matter of fact, LFSR sequences do support leapfrog and block splitting
very well.  Block splitting means basically jumping ahead in a PRN sequence. In
the case of LFSR sequences this can be done quite efficiently. Note, that by
introducing a companion matrix $A$ the linear recurrence (\ref{eq:lfsr}) can
be written as a vector matrix product \cite{LEcuyer:1990:simulation}.
\begin{equation}
  \begin{pmatrix}
    r_{i-(n-1)} \\ \vdots \\ r_{i-1} \\ r_{i}
  \end{pmatrix}=\underbrace{
  \begin{pmatrix}
    0 & 1 & \dots & 0 \\
    \vdots & \vdots & \ddots & \vdots \\
    0 & 0 & \dots & 1 \\
    a_n & a_{n-1} & \dots & a_1 
  \end{pmatrix}}_{\displaystyle A}
  \begin{pmatrix}
    r_{i-n} \\ \vdots \\ r_{i-2} \\ r_{i-1}
  \end{pmatrix}
  \bmod m
\end{equation}
From this formula it follows immediately that the $M$-fold successive
iteration of (\ref{eq:lfsr}) may be written as
\begin{equation}
  \begin{pmatrix}
    r_{i-(n-1)} \\ \vdots \\ r_{i-1} \\ r_{i}
  \end{pmatrix}=
  A^M
  \begin{pmatrix}
    r_{i-M-(n-1)} \\ \vdots \\ r_{i-M-1} \\ r_{i-M}
  \end{pmatrix}
  \bmod m\,.
\end{equation}
Matrix exponentiation can be accomplished in $\bigO{n^3\ln M}$ steps via
binary exponentiation (also known as exponentiation by squaring).

Implementing leapfrogging efficiently is less straightforward.
Calculating $t_{j,i}=r_{pi+j}$ via
\begin{equation}
  \begin{pmatrix}
    r_{pi+j-(n-1)} \\ \vdots \\ r_{pi+j-1} \\ r_{pi+j}
  \end{pmatrix}=
  A^p
  \begin{pmatrix}
    r_{p(i-1)+j-(n-1)} \\ \vdots \\ r_{p(i-1)+j-1} \\ r_{p(i-1)+j}
  \end{pmatrix}
  \bmod m
\end{equation}
is no option, because $A^p$ is usually a dense matrix, in which case
calculating a new element from the leapfrog sequence requires $\bigo(n^2)$
operations instead of $\bigo(n)$ operations in the base sequence.

The following theorem assures that the leapfrog subsequences of LFSR sequences
are again LFSR sequences
\cite{Jungnickel:1993:Finite_Fields}. This will provide us with a very
efficient way to generate leapfrog sequences.
\begin{theorem}
  \label{the:leapfrog1}%
  Let $(r)$ be a LFSR sequence based on a primitive polynomial of degree $n$
  with period $m^n-1$ (pseudonoise sequence) over $\F_m$, and let $(t)$ be
  the decimated sequence with lag $p>0$ and offset $j$, e.\,g.,
  \begin{equation}
    \label{eq:def-decimated}
    t_{j,i} = r_{pi+j}\,.
  \end{equation}
  Then $(t_j)$ is a LFSR sequence based on a primitive polynomial of degree
  $n$, too, if and only if $p$ and $m^n-1$ are coprime, e.\,g., greatest
  common divisor $\gcd(m^n-1, p)=1$.  In addition, $(r)$ and $(t_j)$ are not
  just cyclic shifts of each other, except when
  \begin{equation}
    \label{eq:cyclicshift}
    p = m^h 
  \end{equation}
  for some $0\leq h < n$. If $\gcd(m^n-1, p)>1$ the sequence $(t_j)$ is still
  a LFSR sequence, but not a pseudonoise sequence.
\end{theorem}
It is not hard to find prime numbers $m$ such that $m^n-1$ has very few (and
large) prime factors. For such numbers, the leapfrog method yields
pseudonoise sequences for any reasonable number of parallel streams, see also
section~\ref{sec:modulus} and appendix~\ref{sec:Sophie-Germain}.

While Theorem~\ref{the:leapfrog1} ensures that leapfrog sequences are not just
cyclic shifts of the base sequence (unless \eqref{eq:cyclicshift} holds), the
leapfrog sequences are cyclic shifts of \emph{each other}. This is true for general
sequences, not just LFSR sequences. Consider an arbitrary sequence $(r)$ with
period $T$. If $\gcd(T, p)= 1$, all leapfrog sequences $(t_1),(t_2),\dots,
(t_p) $ are cyclic shifts of each other, i.\,e.\ for every pair of leapfrog
sequences $(t_{j_1})$ and $(t_{j_2})$ of a common base sequence $(r)$ with
period $T$ there is a constant $s$, such that $t_{{j_1},i}=t_{{j_2},i+s}$ for
all $i$, and $s$ is at least $\lfloor T/p\rfloor$. Furthermore, if $\gcd(T,
p)=d>1$, the period of each leapfrog sequence equals $T/d$ and there are $d$
classes of leapfrog sequences. Within a class of leapfrog sequences there are
$p/d$ sequences, each sequence is just a cyclic shift of another and the size
of the shift is at least $\lfloor T/p\rfloor$.

Theorem~\ref{the:leapfrog1} tells us that all leapfrog sequences of a LFSR
sequence of degree $n$ can be generated by another LFSR of degree $n$ or less.
The following theorem \cite[Theorem 6.6.1]{Jungnickel:1993:Finite_Fields}
gives us a recipe to calculate the coefficients
$(b_1,b_2,\ldots,b_n)$ of the corresponding leapfrog feedback polynomial:
\begin{theorem}
  \label{the:leapfrog2} 
  Let $(t)$ be a (periodic) LFSR sequence over the field $\F_m$ and $f(x)$ its
  characteristic polynomial of degree $n$. Then the coefficients $(b_{1},
  b_{2}, \dots, b_{n})$ of $f(x)$ can be computed from $2n$ successive
  elements of $(t)$ by solving the linear system
  \begin{equation}
    \label{eq:lin_sys}
    \begin{pmatrix}
      t_{i+n} \\ t_{i+n+1} \\ \vdots \\ t_{i+2n-1} 
    \end{pmatrix}
    =
    \begin{pmatrix}
      t_{i+n-1} & \dots & t_{i+1} & t_{i} \\
      t_{i+n} & \dots & t_{i+2} & t_{i+1} \\
      \vdots & \ddots & \vdots & \vdots \\
      t_{i+2n-2} & \dots & t_{i+n} & t_{i+n-1} 
    \end{pmatrix}
    \begin{pmatrix}
      b_{1} \\ b_{2} \\ \vdots \\ b_{n} 
    \end{pmatrix}
  \end{equation}
  over $\F_m$.
\end{theorem}
Starting from the base sequence we determine $2n$ values of the sequence $(t)$
by applying the leapfrog rule. Then we solve \eqref{eq:lin_sys} by Gaussian
elimination to get the characteristic polynomial for a new LFSR generator that
yields the elements of the leapfrog sequence directly with each call.  If the
matrix in \eqref{eq:lin_sys} is singular, the linear system has more than one
solution, and it is sufficient to pick one of them. In this case it is always
possible to generate the leapfrog sequence by a LFSR of degree less than the
degree of the original sequence.

\subsection{Choice of modulus}
\label{sec:modulus}

LFSR sequences over $\F_2$ with sparse feedback polynomials are popular
sources of PRNs
\cite{Kirkpatrick:Stoll:1981:R250,Ziff:1998:Four-tap,Knuth:1998:TAoCPII} and
generators of this type can be found in various software libraries. This is
due to the fact that multiplication in $\F_2$ is trivial, addition reduces to
exclusive-or and the $\bmod$-operation comes for free. As a result, generators
that operate in $\F_2$ are extremely fast.  Unfortunately, these generators
suffer from serious statistical defects
\cite{Ferrenberg:Landau:Wong:1992:Hidden_errors,Grassberger:1993:correlations_in_good_RNGs,Shchur:Heringa:Blote:1997:directed_radom_walk,Ziff:1998:Four-tap}
that can be blamed quite generally on the small size of the underlying field
\cite{Bauke:Mertens:2004:random_coins}, see also section~\ref{sec:Cluster_MC}.
In parallel applications we have the additional drawback, that, if the
leapfrog method is applied to a LFSR sequence with sparse characteristic
polynomial, the new sequence will have a dense polynomial. The computational
complexity of generating values of the LFSR sequence grows from $\bigO{1}$ to
$\bigO{n}$.  Remember that for generators in $\F_2$, $n$ is typically of order
$1000$ or even larger to get a long period $2^n-1$.

The theorems and parallelization techniques we have presented so far do apply
to LFSR sequences over any finite field $\F_m$.  Therefore we are free to
choose the prime modulus $m$.  In order to get maximum entropy on the
macrostate level \cite{Mertens:Bauke:2004:Entropy} $m$ should be as large as
possible.  A good choice is to set $m$ to a value that is of the order of the
largest representable integer of the computer. If the computer deals with
$e$-bit registers, we may write the modulus as $m=2^e-k$, with $k$ reasonably
small. In fact if $k(k+2)\le m$ modular reduction can be done reasonably fast
by a few bit-shifts, additions, and multiplications, see
section~\ref{sec:modular_reduction}.  Furthermore a large modulus allows us to
restrict the degree of the LFSR to rather small values, e.\,g., $n\approx4$,
while the PRNG has a large period and good statistical properties.

In accordance with Theorem~\ref{the:leapfrog1} a leapfrog sequence of a
pseudonoise sequence is a pseudonoise sequence, too, if and only if its period
$m^n-1$ and the lag $p$ are coprime.  For that reason $m^n-1$ should have a
small number of prime factors \cite{LEcuyer:1999:Good_Params}.  It can be
shown that $m^n-1$ has at least three prime factors and if the number of prime
factors does not exceed three, then $m$ is necessarily a Sophie-Germain prime
and $n$ a prime larger than two, see appendix~\ref{sec:Sophie-Germain} for
details.

To sum up, the modulus $m$ of a LFSR sequence should be a Sophie-Germain
prime, such that $m^n-1$ has no more than three prime factors and such that
$m=2^e-k$ and $k(k+2)\le m$ for some integers $e$ and $k$.

\section{Delinearization}
\label{sec:delinear}

\begin{figure}%
  \includegraphics[width=0.8\linewidth]{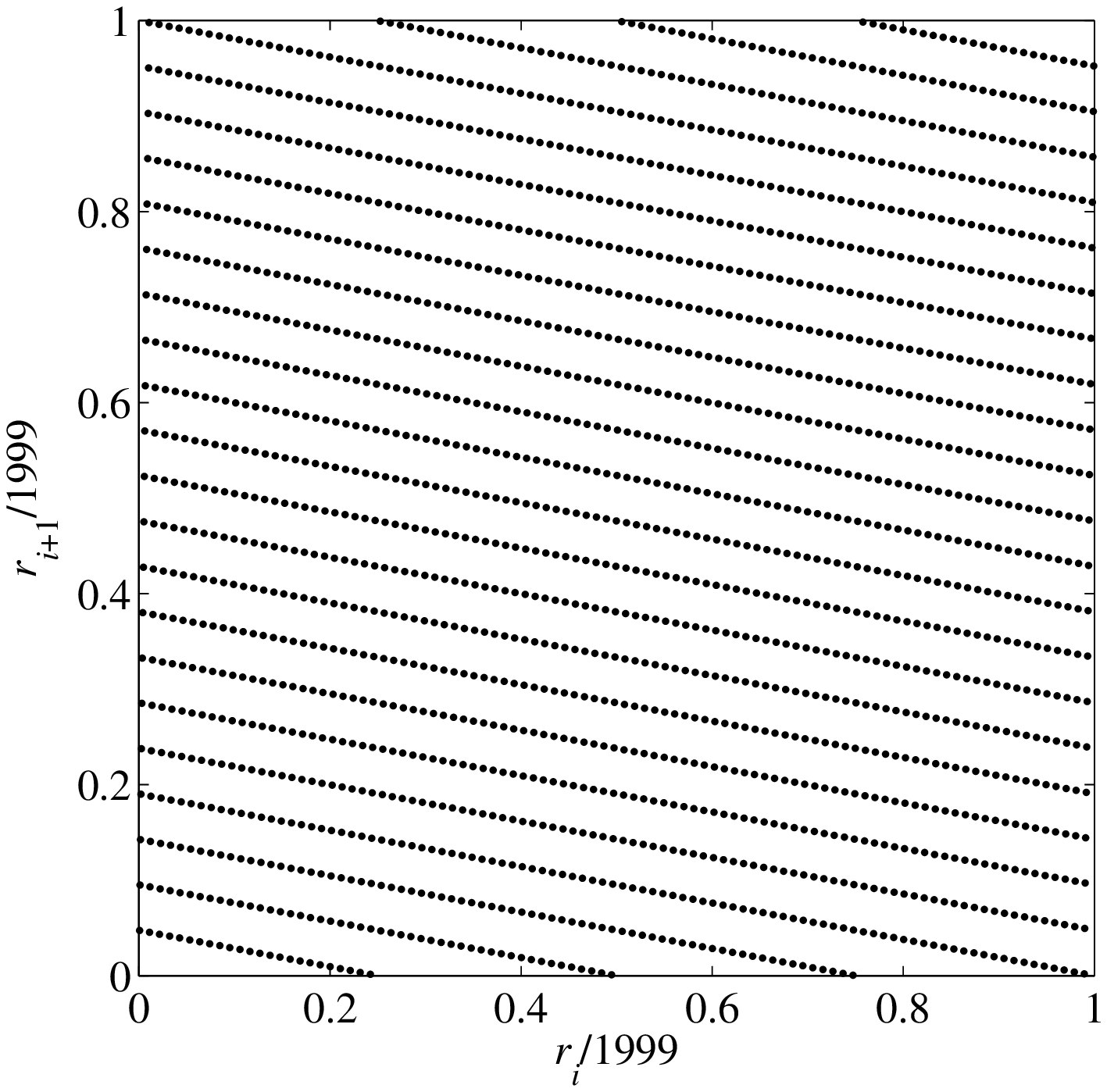}\\[2.5ex plus 1ex minus 1ex]
  \includegraphics[width=0.8\linewidth]{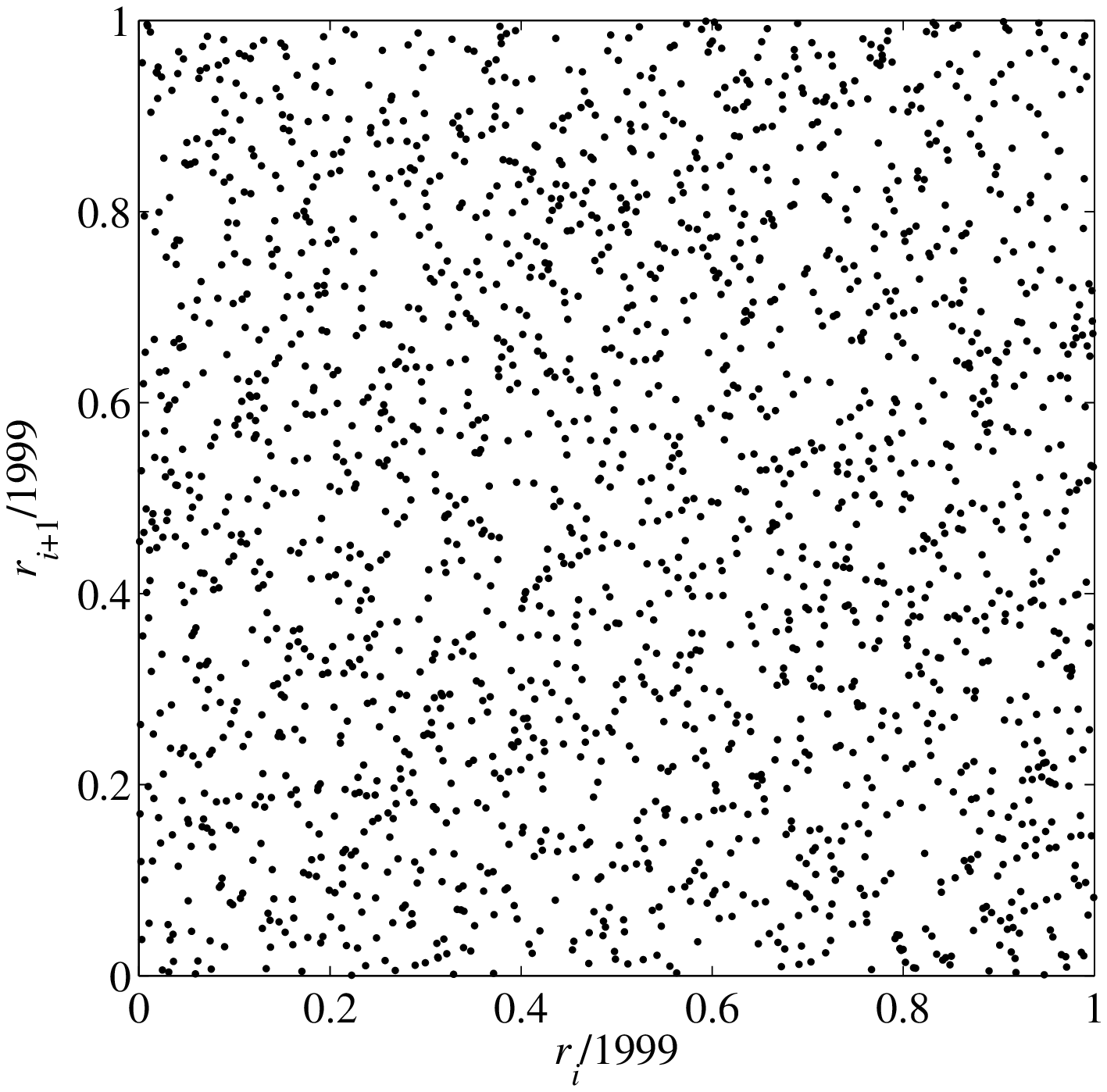}%
  \caption{Exponentiation of a generating element in a prime field is an
    effective way to destroy the linear structures of LFSR sequences. 
    Both pictures show the full period of the generator. Top: 
    $r_i = 95\cdot r_{i-i}\bmod 1999$. Bottom: $r_i = 1099^{q_i}\bmod 1999$ with
    $q_i = 95\cdot q_{i-i}\bmod 1999$.}
  \label{fig:lattice}
\end{figure}

LFSR sequences over prime fields with a large prime modulus seem to be ideally
suited as PRNGs, especially in parallel environments.  They have, however, a
well known weakness.  When used to sample coordinates in $d$-dimensional
space, pseudonoise sequences cover every point for $d < n$, and every point
except $(0,0,\ldots,0)$ for $d=n$. For $d > n$ the set of positions generated
is obviously sparse, and the linearity of the production rule (\ref{eq:lfsr})
leads to the concentration of the sampling points on $n$-dimensional
hyperplanes \cite{Grube:1973:Mehrfach_rekursiv,LEcuyer:1993:Good_MRG}, see
also top of \figurename~\ref{fig:lattice}.  This phenomenon, first noticed by
Marsaglia in 1968 \cite{Marsaglia:1968:planes}, motivates one of the well
known tests for PRNGs, the so-called spectral test
\cite{Knuth:1998:TAoCPII}. The spectral test checks the behavior of a
generator when its outputs are used to form $d$-tuples.  LFSR sequences can
fail the spectral test in dimensions larger than the register size $n$.
Closely related to this mechanism are the observed correlations in other
empirical tests such as the birthday spacings test and the collision test
\cite{LEcuyer:2001:Software_for_uniform_random_number_generation,LEcuyer:Hellekalek:1998:RNGs}.
Nonlinear generators do quite well in all these tests, but compared to LFSR
sequences they have much less nice and \emph{provable} properties and they are
not suited for fair playing parallelization.

To get the best of both worlds we propose a delinearization that preserves
all the nice properties of linear pseudonoise sequences:
\begin{theorem}
  \label{the:yarn}
  Let $(q)$ be a pseudonoise sequence in $\F_m$, and let $g$ be a generating
  element of the multiplicative group $\F^*_m$. Then the sequence $(r)$ with
  \begin{equation}
    \label{eq:def-yarn}
    r_i = 
    \begin{cases}
      g^{q_i}\bmod m & \text{if $q_i > 0$} \\
      0 & \text{if $q_i = 0$}
    \end{cases}
  \end{equation}
  is a pseudonoise sequence, too.
\end{theorem}
The proof of this theorem is trivial: since $g$ is a generator of $\F^*_m$,
the map (\ref{eq:def-yarn}) is bijective. We call delinearized generators
based on Theorem~\ref{the:yarn} YARN generators (yet another random number).

\begin{figure}
  \includegraphics[width=\linewidth]{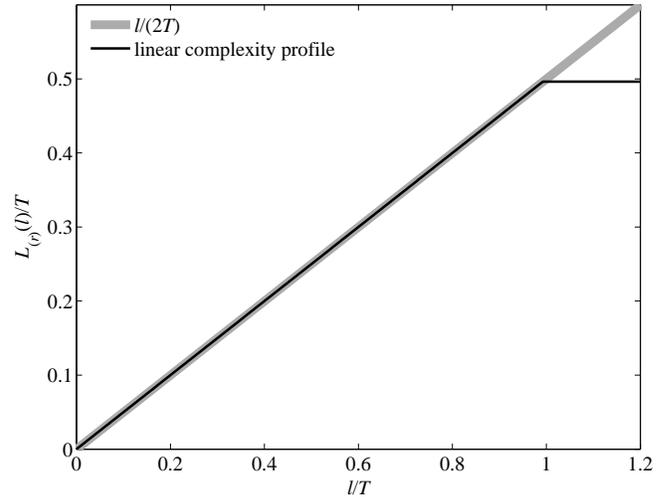}
  \caption{Linear complexity profile $L_{(r)}(l)$ of a YARN sequence, produced
    by the recurrence $q_i = 173\cdot q_{i-1} + 219 \cdot q_{i-2}\bmod 317$
    and $r_i = 151^{q_i} \bmod 317$. The period of this sequence equals
    $T=317^2-1$.}
  \label{fig:linear_complexity_profile}
\end{figure}

The linearity is completely destroyed by the map \eqref{eq:def-yarn}, see
\figurename~\ref{fig:lattice}.  Let $L_{(r)}(l)$ denote the linear complexity
of the subsequence $(r_1,r_2,\dots,r_l)$.  This function is known as the
linear complexity profile of $(r)$. For a truly random sequence it grows on
average like $l/2$.  \Figurename~\ref{fig:linear_complexity_profile} shows the
linear complexity profile $L_{(r)}(l)$ of a typical YARN sequence.  It shows
the same growth rate as a truly random sequence up to the point where more
than 99\,\% of the period has been considered. Sharing the linear complexity
profile with a truly random sequence, we may say that the YARN generator is as
nonlinear as it can get.

\section{Implementation and efficiency}
\label{sec:implementation}

LFSR sequences over prime fields larger than $\F_2$ have been proposed as
PRNGs in the literature
\cite{Grube:1973:Mehrfach_rekursiv,LEcuyer:1990:simulation,Knuth:1998:TAoCPII}.
An efficient implementation of these recurrences requires some care, however.

We assume that all integer arithmetic is done in $w$-bit registers and
$m<2^{w-1}$. Under this condition addition modulo $m$ can be done without
overflow problems.  But multiplying two $(w-1)$-bit integers modulo $m$ is not
straightforward because the intermediate product has $2(w-1)$ significant bits
and cannot be stored in a $w$-bit register.  For the special case
$a_k<\sqrt{m}$ Schrage \cite{Schrage:1979:More_Portable} showed how to
calculate $a_k\cdot r_{i-k}\bmod m$ without overflow. Based on this technique
a portable implementation of LFSR sequences with coefficients $a_k<\sqrt{m}$
is presented in \cite{LEcuyer:Blouin:Couturee:1993:Good_MRGs}. For parallel
PRNGs these methods do not apply because the leapfrog method may yield
coefficients that violate this condition.  Knuth \cite[section
3.2.1.1]{Knuth:1998:TAoCPII} proposed a generalization of Schrage's method for
arbitrary positive factors less than $m$, but this method requires up to
twelve multiplications and divisions and is therefore not very efficient.

The only way to implement (\ref{eq:lfsr}) without additional measures to
circumvent overflow problems is to restrict $m$ to $m<2^{w/2}$.  On machines
with $32$-bit registers, $16$ random bits per number is not enough for some
applications.  Fortunately todays C~compilers provide fast \mbox{$64$-bit}
arithmetic even on \mbox{$32$-bit} CPUs and genuine \mbox{$64$-bit} CPUs
become more and more common.  This allows us to increase $m$ to $32$.

\subsection{Efficient modular reduction}
\label{sec:modular_reduction}

Since the modulo operation in (\ref{eq:lfsr}) is usually slower than other
integer operations such as addition, multiplication, boolean operations or
shifting, it has a significant impact on the total performance of PRNGs based
on LFSR sequences.  If the modulus is a Mersenne prime $m=2^e-1$, however, the
modulo operation can be done using only a few additions, boolean operations
and shift operations \cite{Payne:Rabung:Bogyo:1969:CodingLehmerPRNG}.

A summand $s=a_k\cdot r_{i-k}$ in (\ref{eq:lfsr}) will never exceed
$(m-1)^2=(2^e-2)^2$ and for each positive integer $s\in[0,(2^e-1)^2]$ there is
a unique decomposition of $s$ into
\begin{equation}
  s = r\cdot 2^{e}+q\quad\text{with}\quad 0\le q<2^{e}\,.
\end{equation}
From this decomposition we conclude
\begin{align*}
  s-r\cdot 2^{e} & = q \\
  s-r(2^{e}-1) & = q+r \\
  s \bmod (2^{e}-1) & = q+r \bmod (2^{e}-1) 
\end{align*}
and $r$ and $q$ are bounded from above by
\begin{equation*}
  q < 2^e 
  \quad\text{and}\quad
  r \leq \lfloor(2^e-2)^2/2^e\rfloor < 2^e-2 
\end{equation*}
respectively, and therefore
\begin{equation*}
  q+r < 2^e + 2^e - 2 = 2m \,.
\end{equation*}
So if $m=2^e-1$ and $s\le(m-1)^2$, $x=s\bmod m$ can be calculated solely by
shift operations, boolean operations and addition, viz
\begin{equation}
  \label{eq:mod_2_e_1}
  x = (s \bmod 2^{e}) + \lfloor s/2^{e} \rfloor\,.
\end{equation}
If (\ref{eq:mod_2_e_1}) yields a value $x \ge m$ we simply subtract $m$.

From a computational point of view Mersenne prime moduli are optimal and we
propose to choose the modulus $m=2^{31}-1$.  This is the largest positive
integer that can be represented by a signed $32$-bit integer variable, and it
is also a Mersenne prime. On the other hand our theoretical considerations
favor Sophie-Germain prime moduli, for which (\ref{eq:mod_2_e_1}) does not
apply directly. But one can generalize (\ref{eq:mod_2_e_1}) to moduli $2^e-k$
\cite{Mascagni:Chi:2004:Parallel_LCG_Sophie-Germain}. Again we start from a
decomposition of $s$ into
\begin{equation}
  s = r\cdot 2^{e}+q\quad\text{with}\quad 0\le q<2^{e}\,,
\end{equation}
and conclude
\begin{align*}
  s-r\cdot 2^{e} & = q \\
  s-r(2^{e}-k) & = q+kr \\
  s \bmod (2^{e}-k) & = q+kr \bmod (2^{e}-k) \,.
\end{align*}
The sum $s'=q+kr$ exceeds the modulus at most by a factor $k+1$, because by
applying
\begin{equation*}
  q < 2^e 
  \quad\text{and}\quad
  r \leq \lfloor(2^e-k-1)^2/2^e\rfloor < 2^e-k-1\,,
\end{equation*}
we get the bound
\begin{equation*}
  q+kr < 2^e + k(2^e-k-1) = (k+1) m \,.
\end{equation*}
In addition, by the decomposition of $s'=q+kr$
\begin{equation*}
  s' = r'\cdot 2^{e}+q'\quad\text{with}\quad 0\le q'<2^{e}\,,
\end{equation*}
it follows
\begin{equation*}
  s \bmod (2^{e}-k) = s' \bmod (2^{e}-k) = q'+kr' \bmod (2^{e}-k) \,,
\end{equation*}
and this time the bounds
\begin{equation*}
  q' < 2^e 
  \quad\text{and}\quad
  r' \leq \lfloor(k+1)(2^e-k)/2^e\rfloor < k+1
\end{equation*}
and
\begin{equation*}
  q'+kr' < 2^e + k(k+1) = m + k(k+2)
\end{equation*}
hold. Therefore if $m=2^e-k$, $s\le(m-k)^2$ and $k(k+2)\le m$, $x=s\bmod m$
can be calculated solely by shift operations, boolean operations and addition,
viz
\begin{equation}
  \label{eq:mod_2_e_k}
  \begin{split}
    s' & = (s \bmod 2^{e}) + k\lfloor s/2^{e} \rfloor \,\\
    x & = (s' \bmod 2^{e}) + k\lfloor s'/2^{e} \rfloor\,.
  \end{split}
\end{equation}
If (\ref{eq:mod_2_e_k}) yields a value $x \ge m$, a single subtraction of $m$
will complete the modular reduction. To carry out (\ref{eq:mod_2_e_k}) twice
as many operations as for (\ref{eq:mod_2_e_1}) are needed. But
(\ref{eq:mod_2_e_k}) applies for all moduli $m=2^e-k$ with $k(k+2)\le m$. Note
that on systems with big enough word size (such as 64-bit architectures), just
doing the $\bmod$-operation might well be faster than using
\eqref{eq:mod_2_e_k}.

\subsection{Fast delinearization}

YARN generators hide linear structures of LFSR sequences $(q)$ by raising a
generating element $g$ to the power $g^{q_i}\bmod m$. This can be done
efficiently by binary exponentiation, which takes $\bigO{\log m}$ steps. But
considering LFSR sequences with only a few feedback taps ($n\leq 6$) and
$m\approx2^{31}$ even fast exponentiation is significantly more expensive than
a single iteration of (\ref{eq:lfsr}). Therefore we propose to implement
exponentiation by table lookup. If $m$ is a $2e'$-bit number we apply the
decomposition
\begin{equation}
  \begin{split}
    q_i & = q_{i,1}\cdot 2^{e'}+q_{i,0}\quad\text{with} \\
    q_{i,1} & = \lfloor q_i/2^{e'}\rfloor\,,\quad q_{i,0} = q_i \bmod 2^{e'} \,,
  \end{split}
\end{equation}
and use the identity
\begin{equation}
  \label{eq:pow_mod}
  r_i = g^{q_i} \bmod m = (g^{2^{e'}})^{q_{i,1}}\cdot g^{q_{i,0}}\bmod m
\end{equation}
to calculate $g^{q_i} \bmod m$ by two table lookups and one multiplication
modulo $m$. If $m<2^{31}$ the tables for $(g^{2^{e'}})^{q_{i,1}}\bmod m$ and
$g^{q_{i,0}}\bmod m$ have $2^{16}$ and $2^{15}$ entries, respectively. These
384~kbytes of data easily fit into the cache of modern CPUs.

\begin{table}[t]
  \caption{Generators of the TRNG library. All PRNGs denoted by
    trng::mrg$n$\{s\} are pseudonoise sequences over $\F_m$ with $n$ feedback
    terms, trng::yarn$n$\{s\} denotes its delinearized counterpart. The prime
    modulus $2^{31}-1$ is a Mersenne prime, whereas all other moduli are
    Sophie-Germain primes.} 
  \label{tab:generators}
  \begin{ruledtabular}
    \begin{tabular}{@{}lll@{}}
      name & $m$ & period \\
      \cline{1-1}\cline{2-2}\cline{3-3}
      trng::mrg2, trng::yarn2   & $2^{31}-1$ & $\approx2^{62}$ \\
      trng::mrg3, trng::yarn3   & $2^{31}-1$ & $\approx2^{93}$ \\
      trng::mrg3s, trng::yarn3s & $2^{31}-21\,069$ & $\approx2^{93}$ \\
      trng::mrg4, trng::yarn4   & $2^{31}-1$ & $\approx2^{124}$ \\
      trng::mrg5, trng::yarn5   & $2^{31}-1$ & $\approx2^{155}$ \\
      trng::mrg5s, trng::yarn5s & $2^{31}-22\,641$ & $\approx2^{155}$ \\
    \end{tabular}
  \end{ruledtabular}
\end{table}

\begin{table}[t]
  \caption{Performance of various PRNGs from the TRNG library version 4.0
    \cite{TRNG} and the GNU Scientific Library (GSL) version~1.8
    \cite{GSL}. The test program was compiled using the Intel C++ compiler
    version~9.1 with high optimization (option \texttt{-O3}) and was executed
    on a Pentium~IV 3\,GHz.}%
  \begin{ruledtabular}
    \begin{tabular}{@{}lrlr@{}}
      \multicolumn{2}{c}{TRNG} &  
      \multicolumn{2}{c}{GSL}\\
      \cline{1-2}\cline{3-4}
      generator & PRNs per sec. & generator & PRNs per sec. \\
      \cline{1-1}\cline{2-2}\cline{3-3}\cline{4-4}
      trng::mrg2   &  $48\cdot10^6$ &       
      mt19937\footnote{Mersenne Twister \cite{Matsumoto:Nishimura:1998:Mersenne_Twister}}
      & $41\cdot10^6$ \\
      trng::mrg3   &  $43\cdot10^6$ &
      r250\footnote{shift-register sequence over $\F_2$, $r_i = r_{i-103} \oplus r_{i-250}$ \cite{Kirkpatrick:Stoll:1981:R250}, see also section~\ref{sec:Cluster_MC}} & $85\cdot10^6$ \\
      trng::mrg3s  &  $36\cdot10^6$ &
      gfsr4\footnote{shift-register sequence over $\F_2$, $r_i = r_{i-471} \oplus r_{i-1586} \oplus r_{i-6988} \oplus r_{i-9689}$ \cite{Ziff:1998:Four-tap}} & $77\cdot10^6$ \\
      trng::mrg4   &  $38\cdot10^6$ & 
      ran2\footnote{Numerical Recipes generator ran2 \cite{Press:etal:1992:NRC}} & $27\cdot10^6$ \\
      trng::mrg5   &  $38\cdot10^6$ &
      ranlux\footnote{RANLUX generator with default luxury level \cite{Luscher:1994:RANLUX}} & $8\cdot10^6$ \\
      trng::mrg5s  &  $23\cdot10^6$ &
      ranlux389\footnote{RANLUX generator with highest luxury level \cite{Luscher:1994:RANLUX}} & $5\cdot10^6$ \\
      trng::yarn2  &  $26\cdot10^6$ \\
      trng::yarn3  &  $23\cdot10^6$ \\
      trng::yarn3s &  $16\cdot10^6$ \\
      trng::yarn4  &  $20\cdot10^6$ \\
      trng::yarn5  &  $22\cdot10^6$ \\
      trng::yarn5s &  $13\cdot10^6$ \\
    \end{tabular}
  \end{ruledtabular}
  \label{tab:performance}
\end{table}

\subsection{TRNG}

LFSR sequences and their nonlinear counterparts have been implemented in the
TRNG software package, a portable and highly optimized library of
parallelizable PRNGs.  Parallelization by block splitting as well as by
leapfrogging are supported by all generators. TRNG is publicly available for
download \cite{TRNG}.  Its design is based on a proposal for the next revision
of the ISO C++ standard
\cite{Brown:Fischler:Kowalkowski:Paterno:2006:Comprehensive_Proposal}.  TRNG
uses \mbox{$64$-bit-arithmetic}, fast modular reduction (\ref{eq:mod_2_e_1})
and (\ref{eq:mod_2_e_k}) and exponentiation by table lookup (\ref{eq:pow_mod})
to implement PRNGs based on LFSR sequences over prime fields, with Mersenne or
Sophie-Germain prime modulus.  \Tablename~\ref{tab:generators} describes the
generators of the software package.

\Tablename~\ref{tab:performance} shows some benchmark results.  Apparently the
performance of both types of PRNGs competes well with popular PRNGs such as
the Mersenne Twister or RANLUX. Absolute as well as relative timings in
\tablename~\ref{tab:performance} depend on compiler and CPU architecture, but
the table gives a rough performance measure of our PRNGs.

\section{Quality}
\label{sec:quality}

It is not possible to \emph{prove} that a given PRNG will work well in any
simulation, but one can subject a PRNG to a battery of tests that mimic
typical applications. One distinguishes empirical and theoretical test
procedures.  Empirical tests take a finite sequence of PRNs and compute
certain statistics to judge the generator as ``random'' or not. While
empirical tests focus only on the statistical properties of a finite stream of
PRNs and ignore all the details of the underlying PRNG algorithm, theoretical
tests analyze the PRNG algorithm itself by number-theoretic methods and
establish \emph{a priori} characteristics of the PRN sequence. These \emph{a
  priori} characteristics may be used to choose good parameter sets for
certain classes of PRNGs. The parameters of the LFSR sequences used in TRNG
for example were found by extensive computer search
\cite{LEcuyer:1993:Good_MRG} to give good results in the spectral test
\cite{Knuth:1998:TAoCPII}.

In this section we apply statistical tests to the generators shown in
\tablename~\ref{tab:generators} to investigate their statistical properties
and those of its leapfrog substreams.  The test procedures are motivated by
problems that occur frequently in physics, and that are known as sensitive
tests for PRNGs, namely the cluster Monte Carlo simulation of the
two-dimensional Ising model and random walks. Furthermore we analyze the
effect of the exponentiation step of the YARN generator in the birthday
spacings test. Results of additional empirical tests are presented in the
documentation of the software library TRNG \cite{TRNG}.

\subsection{Cluster monte carlo simulations}
\label{sec:Cluster_MC}

In \cite{Ferrenberg:Landau:Wong:1992:Hidden_errors} it is reported, that some
``high quality'' generators produce systematically wrong results in a
simulation of the two-dimensional Ising model at the critical temperature of
the infinite system by the Wolff cluster flipping algorithm
\cite{Wolff:1989:Collective_MC}.  Deviations are due to the bad mixing
properties of the generators (low macrostate entropy)
\cite{Mertens:Bauke:2004:Entropy}. An infamous example for such a bad PRNG is
the r250 generator \cite{Kirkpatrick:Stoll:1981:R250}. It combines $e$ LFSR
sequences over $\F_2$ to produce a stream of $e$-bit integers via
\begin{equation}
  \label{eq:r250}
  r_i = r_{i-103} \oplus r_{i-250}\,,
\end{equation}
where $\oplus$ denotes the bitwise exclusive-or operation (addition in
$\F_2$).

\begin{figure}
  \includegraphics[width=\linewidth]{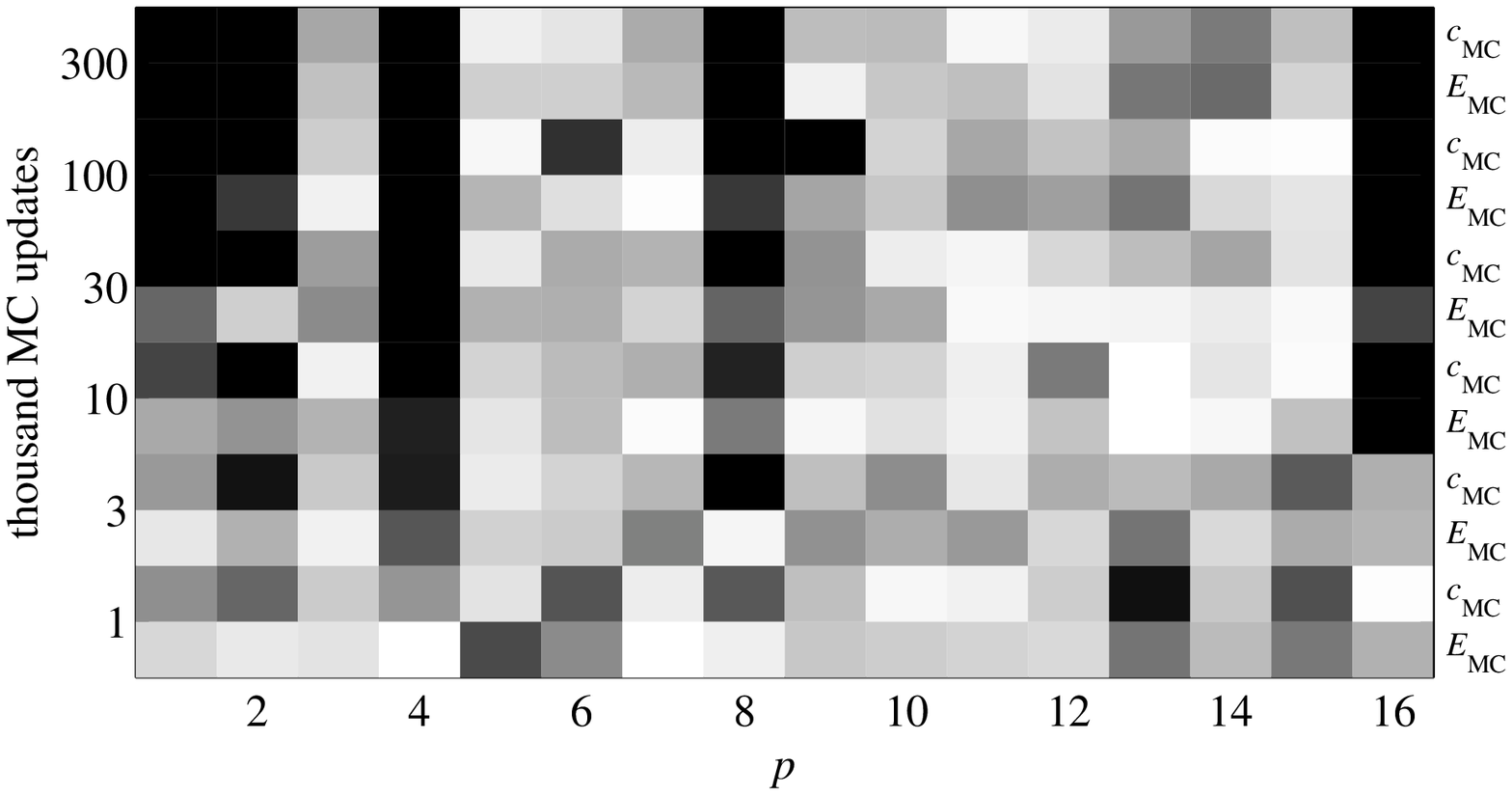}\\[2ex plus 1ex]
  \includegraphics[width=\linewidth]{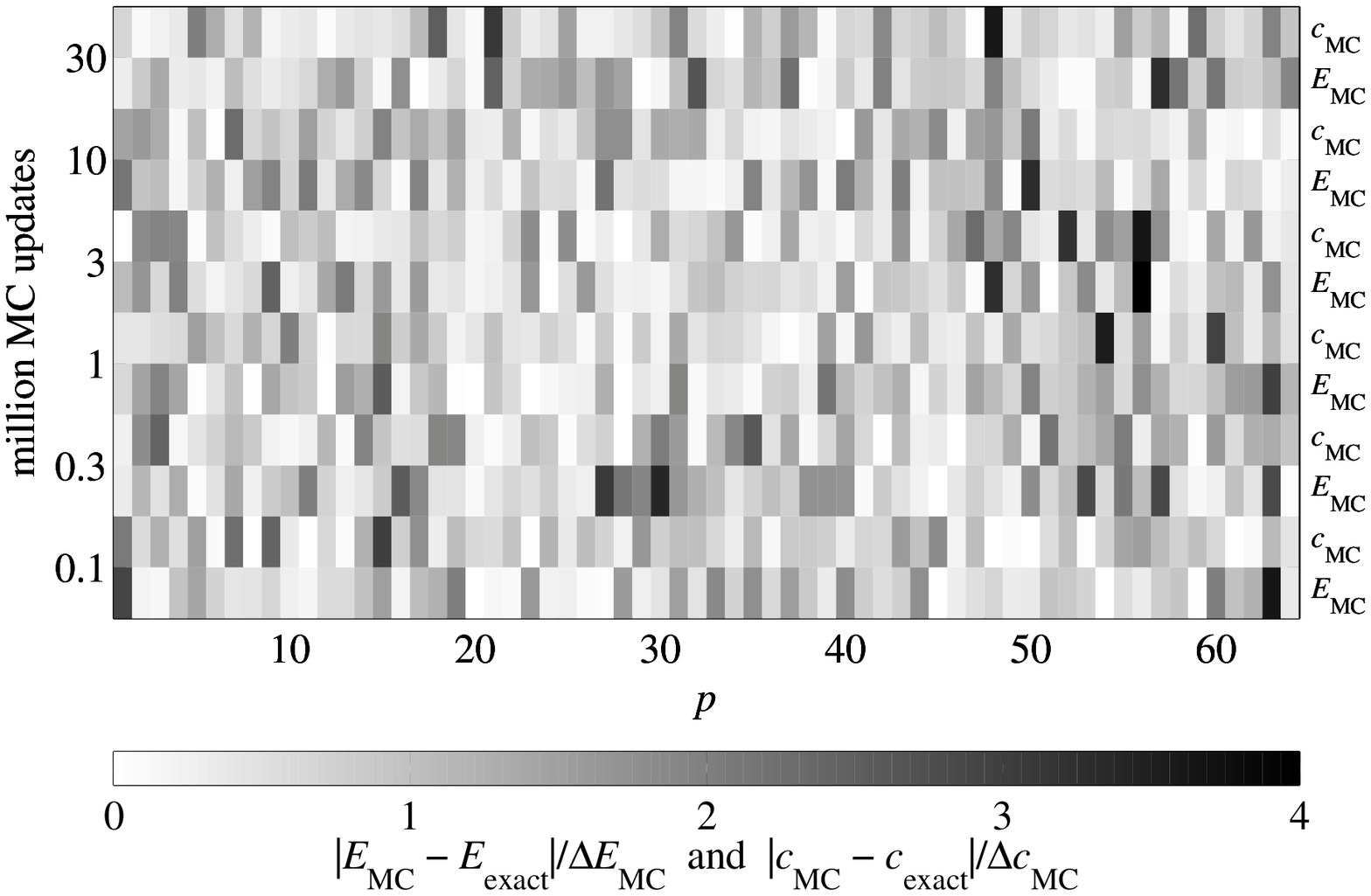}
  \caption{Results of Monte Carlo simulations of the two-dimensional Ising
    model at the critical temperature using the Wolff cluster flipping
    algorithm and the generator r250 (top) and trng::mrg3s (bottom). Only each
    $p$th random number of the sequence was used during the simulations. See
    the text for details.}
  \label{fig:wolff}
\end{figure}

\Figurename~\ref{fig:wolff} shows the results of cluster Monte Carlo
simulations of the Ising model on a $16\times16$ square lattice with cyclic
boundary conditions for the generator r250 and its leapfrog substreams. Each
simulation was done ten times at the critical temperature applying the Wolff
cluster algorithm. The mean of the internal energy $E_{\mathrm{MC}}$, the
specific heat $c_{\mathrm{MC}}$, and the empirical standard deviation of both
quantities have been measured.  The gray scale in \figurename~\ref{fig:wolff}
codes the deviation from the exact values
\cite{Ferdinand:Fisher:1969:Ising_Finite_Lattice}.  Black bars indicate
deviations larger than four standard deviations from the exact values. The
original r250 sequence and its leapfrog sequences yield bad results if the
number of processes is a power of two.  Since the statistical error decreases
as the number of Monte Carlo updates increases, these systematic deviations
become more and more significant and a PRNG with bad statistical properties
will reveal its defects.  \Figurename~\ref{fig:wolff} visualizes the inverse
relation between quantity and quality of streams of PRNs mentioned in the
introduction.

The apparent dependency of the quality on the leapfrog parameter $p$ is easily
understood: If $p$ is a power of two, the leapfrog sequence is just a shifted
version of the original sequence
\cite{Golomb:1982:Shift_Register_Sequences,Ziff:1998:Four-tap}. On the other
hand, if $p$ is not a power of two, the leapfrog sequence corresponds to a
LFSR sequence with a denser characteristic polynomial. For $p=3$, e.\,g., the
recursion reads
\begin{align*}
  r_i & = r_{i-103} \oplus r_{i-152} \oplus r_{i-201} \oplus r_{i-250}\,,\\
  \intertext{and for $p=7$,}
  r_i & = r_{i-103} \oplus r_{i-124} \oplus r_{i-145} \oplus r_{i-166}\oplus 
  r_{i-187} \oplus\null \\
  & \qquad 
  r_{i-208} \oplus r_{i-229} \oplus r_{i-250} \,,
\end{align*}
respectively. The quality of a LFSR generator usually increases with the
number of nonzero coefficients in its characteristic polynomial 
\cite{Wang:Compagner:93,Tezuka:book,Matsumoto:Kurita:96}, a phenomenon that
can be easily explained theoretically
\cite{Bauke:Mertens:2004:random_coins}, and that also holds for
LFSR sequences over general prime fields \cite{Mertens:Bauke:2004:Entropy}. 

While LFSR sequences over $\F_2$ with sparse characteristic polynomial are
known to fail in cluster Monte Carlo simulations, LFSR sequences over large
prime fields with dense characteristic polynomial perform very well in these
tests. This can be seen in the lower half of \figurename~\ref{fig:wolff}
for the trng::mrg3s generator; other generators from
\tablename~\ref{tab:generators} perform likewise.

\subsection{Random walks}

The simulation of the two-dimensional Ising model in the previous section
measured the quality within distinct substreams. Potential interstream
correlations are not taken into account.  In~\cite{Vattulainen:1999:Framework}
some tests are proposed that are designed to detect correlations between
substreams. One of them is the so-called $S_N$~test.

\begin{figure}
  \includegraphics[width=\linewidth]{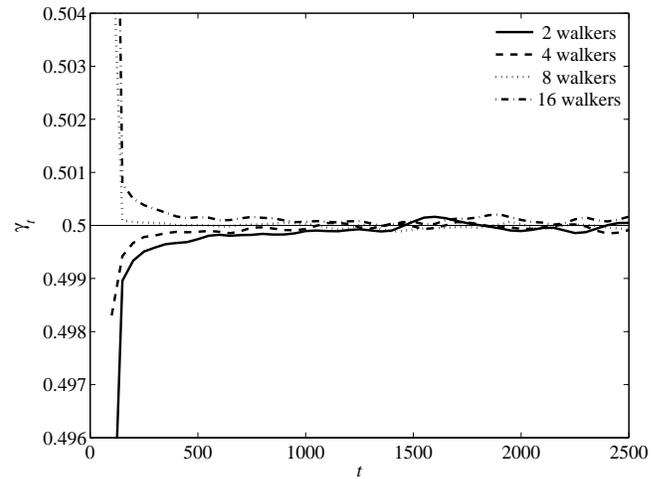}
  \caption{The number of sites visited by $N$ random walkers after $t$ time
    steps is $S_N(t)\sim t^\gamma$. The plot shows the exponent $\gamma$
    against time for generator trng::yarn3.}
  \label{fig:sntest_YARN3}
\end{figure}

For the $S_N$~test $N$ random walkers are considered. They move simultaneously
and independently on the one-dimen\-sional lattice, i.\,e., on the set of all
integers.  At each time step $t$ each walker moves one step to the left or to
the right with equal probability.  For large times $t$ the expected number
$S_N(t)$ of sites that have been visited by at least one walker reaches the
asymptotic form $S_N(t)\sim t^\gamma$ with $\gamma=1/2$. In the $S_N$~test the
exponent $\gamma$ is measured and compared to the asymptotic value
$\gamma=1/2$.

\begin{figure}[t]
  \includegraphics[width=\linewidth]{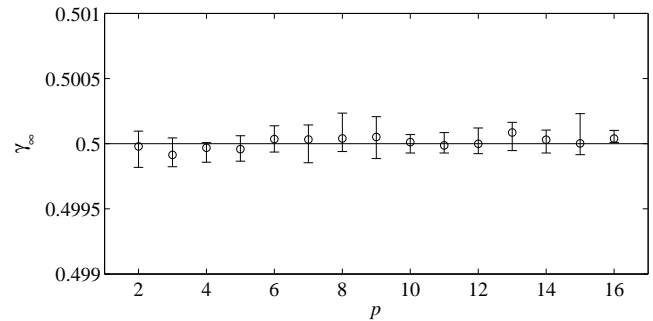}
  \caption{Results of the $S_N$~test for generator trng::mrg5; corresponding
    results for other generators in \tablename~\ref{tab:generators} look
    similar. See the text for details.}
  \label{fig:sntest}
\end{figure}

Let $p$ be the number of parallel streams or walkers. The motion of each
random walker is determined by a leapfrog stream $t_{i,j}$ based on the
generators from \tablename~\ref{tab:generators}.  The exponent $\gamma$ is
determined for up to $16$ simultaneous random walkers as a function of time
$t$. After about $1000$ time steps the exponent has converged and fluctuates
around its mean value.

\Figurename~\ref{fig:sntest_YARN3} shows some typical plots of the exponent
$\gamma$ against time $t$ for generator trng::yarn3. The exponent $\gamma$ was
determined by averaging over $10^8$ random walks. Results for the other
generators in \tablename~\ref{tab:generators} look similar and are omitted
here.  \Figurename~\ref{fig:sntest} presents the asymptotic values of $\gamma$
as a function of the number of walkers for generator trng::mrg5.  The mean
value of $\gamma$ averaged over the interval $1500<t<2500$ is plotted and the
error bars indicate the maximum and the minimum value of $\gamma$ in
$1500<t<2500$.  Almost all numerical results are in good agreement with the
theoretical prediction $\gamma=1/2$ and for other generators we get
qualitatively the same results. Therefore we conclude that there are no
interstream correlations detectable by the $S_N$~test.

\subsection{Birthday spacings test}

The birthday spacings test is an empirical test procedure that was proposed by
Marsaglia \cite{Marsaglia:1985:Current_View,Knuth:1998:TAoCPII}. It is
specifically designed to detect the hyperplane structures of LFSR sequences
\cite{LEcuyer:2001:Software_for_uniform_random_number_generation}.

For the birthday spacings test $N$ days $0\le d_i<M$ are uniformly and
randomly chosen in a year of $M$ days. After sorting these independent
identically distributed random numbers into nondecreasing order $b_{(1)},
b_{(2)}, \dots, b_{(N)}$, $N$ spacings
\begin{align*}
  s_1 & = b_{(2)}-b_{(1)}\,,\\
  s_2 & = b_{(3)}-b_{(2)}\,,\\
  & \Vdots \\
  s_N & = b_{(1)}+M-b_{(N)}
\end{align*} 
are defined. The birthday spacings test examines the distribution of these
spacings by sorting them into nondecreasing order $s_{(1)}, s_{(2)}, \dots,
s_{(N)}$ and counts the number $R$ of equal spacings. More precisely, $R$ is
the number of indices $j$ such that $1\le j<N$ and $s_{(j)}=s_{(j+1)}$. The
test statistics $R$ is a random number with distribution
\begin{equation}
  \label{eq:spacings_dist}
  p_R(r) = \frac{[N^3/(4M)]^r}{\e^{N^3/(4M)}r!} + \bigO{\frac{1}{N}} \,.
\end{equation}
For $N\to\infty$ and $M\to\infty$ such that
$N^3/(4M)\to\lambda=\mathrm{const.}$, $p_R(r)$ converges to a Poisson
distribution with mean $\lambda$. After repeating the birthday
spacings test several times one can apply a chi-square test to compare the
empirical distribution of $R$ to the correct distribution
(\ref{eq:spacings_dist}). If the $p$-value of the chi-square test is very
close to one or zero, the birthday spacings test has to be considered failed
\cite{Knuth:1998:TAoCPII}.

In order to make the birthday spacings test sensitive to the hyperplane
structure of linear sequences, the birthdays are arranged in a
$d$-dimen\-sional cube of length $l$, i.\,e., $M=l^d$. Each day $b_i$ is
determined by a $d$-tuple of consecutive PRNs $(r_{di}, r_{di+1}, \dots,
r_{di+d-1})$,
\begin{equation}
  \label{eq:flat_birthdays}
  b_i = \sum_{j=0}^{d-1} \left\lfloor
    \frac{l\cdot r_{di+j}}{m} 
  \right\rfloor l^j \,.
\end{equation}
This bijective mapping transforms points in a $d$-di\-men\-sional space $[0,
1,\dots,l-1]^d$ to the linear space $[0,1,\dots,M-1]$ and points on regular
hyperplanes are transformed into regular spacings between points in
$[0,1,\dots,M-1]$.

\begin{figure}
  \includegraphics[width=\columnwidth]{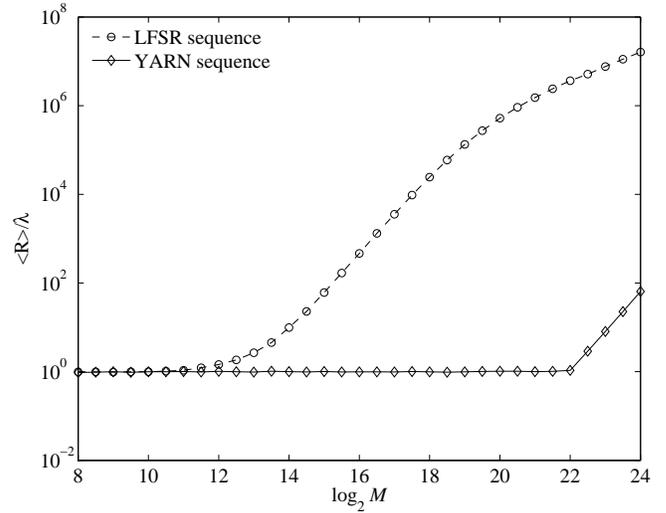}\\[1ex]
  \includegraphics[width=0.9\columnwidth]{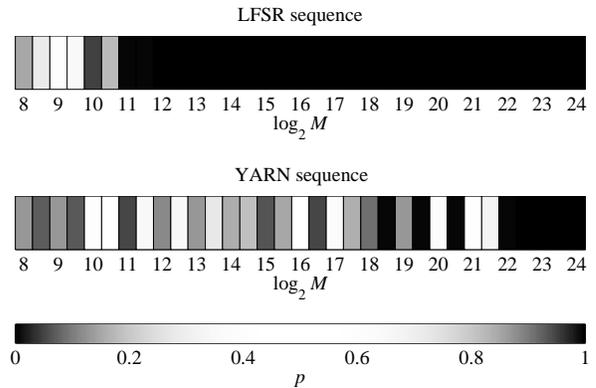}
  \caption{Due to their lattice structure LFSR generators fail the birthday
    spacings test. The nonlinear mapping of YARN sequences is an effective
    technique to destroy these lattice structures. Top: mean number $\langle
    R\rangle$ of equal spacings between $M$ randomly chosen birthdays. For a
    good PRNG $\langle R\rangle/\lambda=1$ is expected. Bottom: gray coded
    $p$-value of a chi-square test for the distribution of $R$.}
  \label{fig:birthdays_spacings}
\end{figure}

To demonstrate the failure of LFSR sequences in the birthday spacings test we
applied the test to the LFSR sequence
\begin{equation}
  \label{eq:LFSR_bst}
  r_i=17384\cdot r_{i-1}+12391\cdot r_{i-2} \mod 65\,521
\end{equation}
and its YARN counterpart
\begin{equation}
  \label{eq:YARN_bst}
  \begin{split}
    r_i & = 
    \begin{cases}
      20009^{q_i}\bmod 65\,521 & \text{if $q_i > 0$} \\
      0 & \text{if $q_i = 0$}
    \end{cases}
    \qquad\text{with}\\
    q_i & = 17384\cdot q_{i-1}+12391\cdot q_{i-2} \bmod 65\,521 
  \end{split}
\end{equation}
with the test parameters $\lambda=N^3/(4M)\approx1$ and $d=6$. Beyond a
certain value of $M$ the LFSR sequence starts to fail the birthday spacings
test due to its hyperplane structure, see
\figurename~\ref{fig:birthdays_spacings}.  The hyperplane structure of LFSR
sequences give rise to birthday spacings that are much more regular than
randomly chosen birthdays. As a consequence we observe a value $R$ that is too
large, see the top of \figurename~\ref{fig:birthdays_spacings}.

The nonlinear transformation (\ref{eq:def-yarn}) in the YARN sequences
destroys the hyperplane structure.  In fact, even those YARN sequences whose
underlying linear part fails the birthday spacings test, passes the same test
with flying colors (\figurename~\ref{fig:birthdays_spacings}).  YARN
generators start to fail the birthday spacings test only for trivial reasons,
i.\,e., when $M$ gets close to the period of the generator.  For the
experiment shown in \figurename~\ref{fig:birthdays_spacings} we averaged over
$5\,000$ realizations of the $M$ birthdays, and each birthday was determined
by $d=6$ random numbers. Then we expect that the YARN generator starts to fail
if $M\cdot 6 \cdot 5\,000 \gtrapprox 65\,521^2-1$, i.\,e.,
$M\gtrapprox2^{17}$.

Of course the period of the PRNGs (\ref{eq:LFSR_bst}) and (\ref{eq:YARN_bst})
is artificially small and the birthday spacings test was carried out with
these particular PRNGs to make the effect of hyperplane structures and its
elimination by delinearization as explicit as possible. ``Industrially sized''
LFSR and YARN generators like those in \tablename~\ref{tab:generators} share
the same structural features, however.  So the results for PRNGs
(\ref{eq:LFSR_bst}) and (\ref{eq:YARN_bst}) can be assumed to present generic
results for LFSR and YARN sequences, respectively.

\section{Conclusions}

We have reviewed the problem of generating pseudorandom numbers in parallel
environments. Theoretical and practical considerations suggest to focus on
linear recurrences in prime number fields, also known as LFSR sequences.
These sequences can be efficiently parallelized by block splitting and
leapfrogging, and both methods enable Monte Carlo simulations to play fair,
i.\,e., to yield results that are independent of the degree of
parallelization. We also discussed theoretical and practical criteria for the
choice of parameters in LFSR sequences.  We then introduced YARN sequences,
which are derived from LFSR sequences by a bijective nonlinear mapping. A YARN
sequence inherits the pseudonoise properties from its underlying LFSR
sequence, but its linear complexity is that of a true random sequence. YARN
sequences share all the advantages of LFSR sequences, but they pass all tests
that LFSR sequences tend to fail due to their low linear complexity.

All our results have been incorporated into TRNG, a publicly available
software library of portable, parallelizable pseudorandom number generators.
TRNG complies with the proposal for the next revision of the ISO C++ standard
\cite{Brown:Fischler:Kowalkowski:Paterno:2006:Comprehensive_Proposal}.

\acknowledgments This work was sponsored by the European Community's FP6
Information Society Technologies programme under contract IST-001935,
EVERGROW, and by the German Science Council DFG under grant \mbox{ME2044/1-1}.

\appendix

\section{Sophie-Germain prime moduli}
\label{sec:Sophie-Germain}

The maximal period of a LFSR sequence (\ref{eq:lfsr}) over the prime field
$\F_m$ is given by $m^n-1$.  We are looking for primes $m$ such that $m^n-1$
has for a fixed $n$ a small number of prime factors.

For $m>2$ the factor $m-1$ is even and the smallest possible number of prime 
factors of $m^n-1$ is three,
\begin{equation}
  m^n-1 = 2\cdot\frac{m-1}{2}\cdot(1+m+m^2+\ldots+m^{m-1})\,.
\end{equation}
If $(m-1)/2$ is also prime, $m$ is called a Sophie-Germain prime or safe
prime.  For $m=2$ there exist values for $n$ such that $2^n-1$ is also prime.
primes of form $2^n-1$ are called Mersenne primes, e.\,g., $2^2-1$ and
$2^{86\,243}-1$ are prime.

If $n$ is composite and $m>2$ the period $T$ is a product of at least four
prime factors,
\begin{equation}
  \begin{split}
    m^{kl}-1 & = (m^k-1)\cdot[(m^k)^{l-1}+(m^k)^{l-2}+\ldots+1] \\
    & = (m-1)\cdot(m^{k-1}+m^{k-2}+\ldots+1)\cdot\null \\
    &\qquad [(m^k)^{l-1}+(m^k)^{l-2}+\ldots+1] \,.
  \end{split}
\end{equation}
Actually, the number of prime factors for composite $n$ is even larger. The
factorization of $m^n-1$ is related to the factorization of the polynomial
\begin{equation}
  f_n(x) = x^n-1
\end{equation}
over $\Z$. This polynomial can be factored into cyclotomic polynomials
$\Phi_k(x)$ \cite{Wan:book}, 
\begin{equation}
  \label{eq:cyclotomic_polynomial_prod}
  f_n(x) = x^n-1 = \prod_{d|n}\Phi_d(x)\,.
\end{equation}
The coefficients of the cyclotomic polynomials are all in $\Z$; so the number
of prime factors of the factorization of $m^n-1$ is bounded from below by the
number of cyclotomic polynomials in the factorization of $x^n-1$. This number
equals the number of natural numbers that divide $n$. If $n$ is prime then
$f_n(x)$ is just the product $\Phi_1(x)\cdot\Phi_n(x)$.

\begin{table}[t]
  \caption{A collection of Sophie-Germain primes $m$ for which $m^n-1$ has a
    minimal number of prime factors.}
  \label{tab:primes}
  \begin{ruledtabular}
    \begin{tabular}{@{}llll@{}}
      $n$ & $m$ & $n$ & $m$ \\
      \cline{1-1}\cline{2-2}\cline{3-3}\cline{4-4}
      $1$ & $2^{31}-525$  & $5$ & $2^{31}-46\,365$ \\
      & $2^{31}-69$       & & $2^{31}-22\,641$ \\
      & $2^{63}-5\,781$     & & $2^{63}-594\,981$ \\
      & $2^{63}-4\,569$     & & $2^{63}-19\,581$ \\[1ex]
      $2$ & $2^{31}-37\,485$ & $6$ & $2^{31}-4\,398\,621$ \\
      & $2^{31}-2\,085$     & & $2^{31}-1\,120\,941$ \\
      & $2^{63}-927\,861$   & & $2^{63}-122\,358\,381$ \\
      & $2^{63}-156\,981$   & & $2^{63}-29\,342\,085$ \\[1ex]
      $3$ & $2^{31}-43\,725$ & $7$ & $2^{31}-50\,949$ \\
      & $2^{31}-21\,069$    & & $2^{31}-6\,489$ \\
      & $2^{63}-275\,025$   & & $2^{63}-92\,181$ \\
      & $2^{63}-21\,129$    & & $2^{63}-52\,425$ \\[1ex]
      $4$ & $2^{31}-305\,829$   \\
      & $2^{31}-119\,565$   \\
      & $2^{63}-3\,228\,621$  \\
      & $2^{63}-156\,981$
    \end{tabular}
  \end{ruledtabular}
\end{table}
If $m$ is a prime larger than two, from $\Phi_1(x)=x-1$ it follows that $m^n-1$
can be factorized into at least three factors, namely
\begin{equation}
  m^n-1 = 
  2\cdot\frac{m-1}{2}\cdot\prod_{d>1;\:\:d|n}\Phi_d(m)\,.
\end{equation}
The period of a maximal period LFSR sequence with linear complexity $n$ over
$\F_m$ is a product of exactly three factors, if and only if $m$ is a
Sophie-Germain prime, $n$ is prime, and $\Phi_n(m)$ is prime also. Note,
$\Phi_2(m)=m+1$ is never prime, if $m$ is an odd prime. Let us investigate
some special cases in more detail.

\begin{description}
\item[Case \boldmath$n=2$\,:] The period $m^2-1$ is a product of $2^3\cdot3$
  and at least two other factors. Using the sieve of Eratosthenes modulo~$12$
  it can be shown, that each large enough prime $m$ can be written as
  $m=12\cdot k + c$, where $k$ and $c$ are integers such that $\gcd(12, c)=1$.
  Factoring the period $m^2-1=(12\cdot k + c)^2-1$ we find
  \begin{equation}
    \label{eq:fac_n_2}
    \begin{split}
      (12\cdot k+1)^{2}-1 & = 2^3\cdot3\cdot k\cdot(6k + 1)\\
      (12\cdot k+5)^{2}-1 & = 2^3\cdot3\cdot (2k+1)\cdot(3 k+1)\\
      (12\cdot k+7)^{2}-1 & = 2^3\cdot3\cdot (3k+2)\cdot(2 k+1)\\
      (12\cdot k+11)^{2}-1 & = 2^3\cdot3\cdot (k+1)\cdot(6 k+5)\,.
    \end{split}
  \end{equation}
\item[Case \boldmath$n=4$\,:] The period $m^4-1$ is a product of
  $2^4\cdot3\cdot5$ and at least three other factors. Using the sieve of
  Erastosthenes modulo~$60$ it can be shown, that each large enough prime $m$
  can be written as $m=60\cdot k+c$, where $k$ and $c$ are integers such that
  $\gcd(60, c)=1$.  Factoring the period $m^4-1=(60\cdot k + c)^2-1$ we find
  \begin{equation}
    \label{eq:fac_n_4}
    \begin{split}
      (60\cdot k+1)^{4}-1 & = 
      2^4\cdot3\cdot5\cdot k\cdot(30 k+1)\\
      &\qquad\cdot(1800 k^{2}+60 k+1)\\[1ex]
      (60\cdot k+7)^{4}-1 & = 
      2^4\cdot3\cdot5\cdot(10 k+1)\cdot(15 k+2)\\
      &\qquad\cdot(360 k^{2}+84 k+5)\\[1ex]
      (60\cdot k+11)^{4}-1 & = 
      2^4\cdot3\cdot5\cdot(5 k+1)\cdot (6 k+1)\\
      &\qquad\cdot(1800 k^{2}+660 k+61)\\[1ex]
      (60\cdot k+13)^{4}-1 & = 
      2^4\cdot3\cdot5\cdot(5 k+1)\cdot(30 k+7)\\
      &\qquad\cdot(360 k^{2}+156 k+17)\\[1ex]
      & \qquad\text{and so on \dots}
    \end{split}
  \end{equation}
\item[Case \boldmath$n=6$\,:] This case is similar to the $n=2$ and $n=4$
  cases. Applying the sieve of Erastosthenes modulo $84$ it can be shown that
  $m^6-1$ is a product of $2^3\cdot3^2\cdot7$ and at least four other factors.
\item[Cases \boldmath$n=3$, $n=5$, $n=7$\,:] Here $n$ is prime, and therefore
  the period $m^n-1$ has at least three factors. The number of factors of
  $m^n-1$ will not exceed three, if $m$ is a Sophie-Germain prime and
  $\Phi_n(m)$ is prime, too.
\end{description}

\noindent In \tablename~\ref{tab:primes} we present a collection of
Sophie-Germain primes $m$, for which $m^n-1$ has a minimal number of prime
factors. For $n$ prime an extended table can be found in
\cite{LEcuyer:1999:good}. If $n$ is prime, its factorization can be found by
(\ref{eq:cyclotomic_polynomial_prod}), and if $n=2$ or $n=4$ by
(\ref{eq:fac_n_2}) and (\ref{eq:fac_n_4}), respectively.  All these primes are
good candidates for moduli of LFSR sequences as PRNGs in parallel
applications. Note that the knowledge of the factorization of $m^n-1$ is
essential for an efficient test of the primitivity of characteristic
polynomials.

\bibliography{random}

\end{document}